\documentclass{emulateapj}
\usepackage{epsfig}
\usepackage{lscape}

\submitted{Accepted for Publication in AJ}

\shortauthors{West et al.}
\shorttitle{HI Selected Galaxies in the SDSS. II. Colors}
\begin{document}

\title{HI Selected Galaxies in the Sloan Digital Sky Survey II: The Colors of Gas-Rich Galaxies}

\author{Andrew A. West\altaffilmark{1,2,3}, 
Diego A. Garcia-Appadoo\altaffilmark{4,5},
Julianne J.Dalcanton\altaffilmark{3}, 
Mike J. Disney\altaffilmark{4},
Constance M. Rockosi\altaffilmark{6},
{\v Z}eljko Ivezi{\'c}\altaffilmark{3}}

\altaffiltext{1}{Corresponding author: aaw@mit.edu}
\altaffiltext{2}{MIT Kavli Institute for Astrophysics and Space Research, 77
  Massachusetts Ave, 37-582c, Cambridge, MA 02139-4307}
\altaffiltext{3}{Department of Astronomy, University of Washington, Box 351580,
Seattle, WA 98195}
\altaffiltext{4}{European Southern Observatory, Alonso de Cordova 3107, Casilla 19001, Vitacura, Santiago 19, Chile}
\altaffiltext{5}{Cardiff School of Physics and Astronomy, Cardiff University, Queens Buildings, The Parade, Cardiff, CF24 3AA, UK}
\altaffiltext{6}{UCO/Lick Observatory, Department of Astronomy and Astrophysics, University of California, Santa Cruz, CA 95064}

\begin{abstract} 
  We utilize color information for an HI-selected sample of 195
  galaxies to explore the star formation histories and physical
  conditions that produce the observed colors.  We show that the HI
  selection creates a significant offset towards bluer colors that can
  be explained by enhanced recent bursts of star formation. There is
  also no obvious color bimodality, because the HI selection restricts
  the sample to bluer, actively star forming systems, diminishing the
  importance of the red sequence. Rising star formation rates are
  still required to explain the colors of galaxies bluer than $g-r
  <$ 0.3.  We also demonstrate that the colors of the bluest galaxies
  in our sample are dominated by emission lines and that stellar
  population synthesis models alone (without emission lines) are not
  adequate for reproducing many of the galaxy colors.  These emission
  lines produce large changes in the $r-i$ colors but leave the $g-r$
  color largely unchanged. In addition, we find an increase in the
  dispersion of galaxy colors at low masses that may be the result of
  a change in the star formation process in low-mass galaxies.

\end{abstract}


\section{Introduction}

The star formation history, metallicity, and current star formation
rate all contribute to the observed colors of a galaxy.  Remarkably,
these factors work in concert to yield a fairly well defined locus in
galaxy color-color space (Strateva et al.~2001).  Deviations from this
locus as well as the morphology of the locus itself, can lead to
significant insight into the underlying processes taking place within
galaxies.  Many previous studies have utilized the broad-band colors
of galaxies to investigate the star formation histories and
metallicities of galaxies (e.g. Tinsley 1972; Searle et al.~1973;
Tinsley \& Gunn 1976; Balcells \& Peletier 1994; Roberts \& Haynes
1994; de Jong 1996; Bell \& de Jong 2000; Galaz et al.~2002; Gavazzi
et al.~2002; Bell et al.~2003; MacArthur et al.~2004; Zackrisson,
Bergvall \& Ostlin 2005; Driver et al.~2006; Skibba et al.~2008).

The advent of large surveys such as the Sloan Digital Sky Survey
(SDSS; York et al.~2000) and the Two Micron All Sky Survey (2MASS;
Skrutskie et al.~2006) has created a wealth of uniform data and the
statistical foothold to investigate the bimodality of galaxies (Baldry
et al.~2004; Kauffmann et al.~2003, 2004), luminosity function
(Blanton et al.~2001; Ball et al.~2006), and the average properties of
nearby galaxies (Blanton et al.~2003a, 2003b, 2005; Geha et al.~2006; Maller et
al.~2008; Skibba et al.~2008). While the large optical and infrared
surveys have made large contributions to our understanding of galaxy
evolution, they only trace the stellar component of galaxies and do
not trace other baryonic material such as cold gas.

Galaxies in the local universe span a range of star formation
histories -- from blue, gas-rich, low-surface-brightness (LSB) galaxies
that are slowly turning their gas into stars with low star forming
efficiencies, to red, gas-poor galaxies, that have formed the bulk of
their stars in the past.  Stars dominate the visible light output of
most galaxies, and thus galaxies detected by traditional optical or
infrared imaging have well developed stellar populations.  In
contrast, the natural way to identify gas-rich, less evolved galaxies
is by their 21 cm HI radio emission.  Aside from its importance for
global star formation, a sample of galaxies with both gaseous and
stellar information allows for a more complete census of the local
baryons (a constraint vital to the calibration of n-body simulations;
Governato et al.~2007; Brooks et al.~2009).

Previous studies have combined large HI surveys with optical and infrared samples, namely the Arecibo Duel Beam and Slice Surveys with the Two
Micron All Sky Survey (2MASS; Jarrett et al.~2000; Rosenberg et
al.~2005) and the merging of HIPASS with SuperCOSMOS (Hambly et al.~2001; Hambly, Irwin \& MacGillivary 2001b; Hambly et al.~2001c;
Doyle et al.~2005).  Rosenberg et al.~(2005) were able to
probe the baryonic content of a large sample of galaxies, but were
limited by the shallow depth of 2MASS, which does not have data for
many of the LSB galaxies in the sample.  The HIPASS/SuperCOSMOS sample of Doyle et al.~(2005) contains optical data for more than 3600 HI selected galaxies but also suffers from the shallow depth of the SuperCOSMOS optical data.

Recent studies have combined the Parkes HI Equatorial Survey (ES) with
the SDSS (Disney et al.~2008; Garcia-Appadoo et al.~2009; West et
al.~2009; hereafter W09).  The deep SDSS optical data provides
information about the stellar content for \emph{all} ES HI sources
where the two surveys overlap.  In addition, the uniform, accurate,
and well-calibrated photometry of the HI-selected ES/SDSS sample
allows for a more detailed exploration of the factors affecting galaxy
colors, particularly for galaxies with large reservoirs of gas.  The
optical colors are reasonably sensitive to age, although IR colors are
needed to constrain metallicity (Bell et al.~2003).  There is thus an
unavoidable degeneracy between age and metallicity when using only the
optical colors available with SDSS (Bell \& de Jong 2001; Bell et
al.~2003).  Our sample does not have a complete set of near IR
counterpart data and some of our results will reflect this limitation.

In this paper, we briefly describe the ES/SDSS sample in \S2 and
examine the colors of HI-selected galaxies by comparing them to
stellar population synthesis models (\S3.1) and the colors of
optically selected galaxies (\S3.2).  We also investigate how line
emission affects the broadband colors of gas-rich galaxies (\S3.3). We
demonstrate an increased dispersion in the colors of galaxies at low
masses, and investigate its possible origins (\S3.4).  We summarize
and discuss our results in \S4.

\section{Data}

The HI data for our sample come from the Parkes Equatorial Survey (ES;
Garcia-Appadoo et al.~2009), a blind HI survey of the southern sky
that covers a velocity range from -1280 to 12700 km/s with an RMS
noise of 13 mJy, using the multibeam receiver on the 64m radio
telescope in Parkes, Australia.  The ES, which is described in detail
in an accompanying paper (Garcia-Appadoo et al.~2009), circles the
celestial equator between -6 $< \delta <$ +10 and contains over 1000
sources in 5738 square degrees.  The raw data forms part of the HI
Parkes All Sky Survey (HIPASS; Barnes et al.~2001; Meyer et al.~2004;
Zwaan et al.~2004; Wong et al.~2006), carried out with the same
instrument over the entire sky between -90 $< \delta <$ +25.  However,
the ES fields were searched much earlier (Garcia-Appadoo et al.~2009)
in readiness for comparison with SDSS data.  While the
search techniques were much the same as those of the HIPASS team and rely
heavily on their procedures, the source lists are not identical.  For
example, the completeness limit of the ES list is 30\% fainter than
the HIPASS limit.  This difference is mainly due to our ability to
follow-up and confirm a higher proportion of the fainter sources, a
process that would be impractical with the larger survey.  The
velocity resolution of the ES HI spectra is 18.0 km\ s$^{-1}$ and the
3$\sigma$ HI mass limit of the survey is $10^6\times
D^2_{\rm{Mpc}}M_{\odot}$, assuming a 200 km\ s$^{-1}$ HI galaxy profile.  For
detailed descriptions of the data acquisition, calibration and
reliability see Garcia-Appadoo et al.~(2009) and the HIPASS analysis
contained in Barnes et al.~(2001), Meyer et al.~(2004), and Zwaan et
al.~(2004).

The optical data for this study come from the Sloan Digital Sky Survey
(SDSS; York et al.~2000; Gunn et al.~1998; Fukugita et al.~1996; Hogg
et al.~2001; Smith et al.~2002; Stoughton et al.~2002; Pier et
al.~2003; Ivezi{\'c} et al.~2004; Gunn et all. 2006) Data Release 2
(DR2; Abazajian et al.~2004) sky area.  The DR2 area is 3324 deg$^2$,
about half of which overlaps with the equatorial ES region discussed
above.  Because the SDSS photometric software ({\tt PHOTO}; Lupton et
al.~2002) was not optimized for angularly large galaxies, all of the
photometry presented for the ES/SDSS survey has been re-processed
using the techniques described in W09.  The W09 study describes the
optical sample selection and matching to ES and presents the updated
SDSS photometry for 195 HI-selected galaxies using both a sky
subtraction procedure that is optimized for angularly large galaxies
and a correction for the over-deblending (``shredding'') of nearby
galaxies by the SDSS pipeline photometry.  While more recent SDSS data releases (e.g. DR7; Abazajian et al. 2009) may contain more of the optical counterparts to ES sources, the re-processed photometry is essential for accurate photometric studies and only exists for the DR2 sample (W09). 

To be included in the ES/SDSS sample, each candidate galaxy had to
meet 4 criteria: 1) the ES recessional velocity must agree to within
twice the velocity width (as measured by the W$_{20}$ value) of the
optically derived redshift; 2) there must be no more than 1
detectable, spatially resolved galaxy within the ES beam at the same
redshift; 3) the candidate galaxy must not extend across two or more
SDSS fields; and 4) all galaxies must be at least 1$^{\prime}$ away
from any saturated foreground stars.  For additional details on
catalog matching, see W09.

\section{Results}
\subsection{Modeling the Colors of HI-selected Galaxies}

\begin{figure}
\centering
\plotone{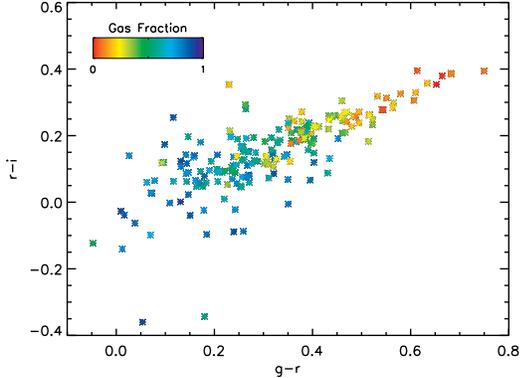}
\caption[$r-i$ vs $g-r$ for HIPASS/SDSS sample galaxies.]{$r-i$ vs
$g-r$ for ES/SDSS sample galaxies.  All photometric corrections
have been applied to these data. Galaxies have been
color coded according to their gas fractions.}
\label{grrigas}
\end{figure}

Figure \ref{grrigas} shows the extinction corrected $r-i$ colors of the galaxies in our HI
selected sample as a function of their $g-r$ color. The galaxies are color-coded according to their gas
fraction.  The gas fraction ($f_{gas}$) is defined as:

\begin{equation}
f_{gas}\equiv\frac{1.4M_{HI}}{1.4M_{HI}+M_{\star}},
\end{equation}

\noindent where M$_{HI}$ is the HI mass and M$_{\star}$ is the stellar
mass determined using color-dependent mass-to-light ratios that have
been corrected for a Kroupa initial mass function (Bell et al.~2003;
Pizagno et al. 2005; see W09).  The factor of 1.4 corrects for the
mass in helium.  We have neglected the molecular gas component, which
is measured to be small in low-mass, LSB galaxies that dominate
our sample (M$_{H_2}$/M$_{HI}<0.1$; Schombert et al. 1990; Matthews et al. 2005).  No molecular gas measurements (CO) exist for any of our sample galaxies. However, recent CO observations of nearby galaxies give some insight to the range molecular gas content in galaxies of similar morphological type. We estimate that the molecular to neutral gas fraction (M$_{H_2}$/M$_{HI}$) may be as high at 0.5 for some of the early-type spirals in the ES/SDSS sample (Leroy et al. 2009). Thus, we may be underestimating the gas fractions in a small number of our galaxies. 

Figure \ref{grrigas} shows several important trends.
First, the HI-selected galaxies occupy a well-defined locus in color,
as seen in other SDSS studies (Strateva et al.~2001; Blanton et
al.~2003a; Baldry et al.~2004; Maller et al.~2008).  Second, the gas
fraction scales with galaxy color such that bluer galaxies have a
larger fraction of their baryonic mass in gas, as has also been
reported for less uniform samples (Kannappan 2004).  Third, the
scatter in color-color space is large for gas-rich galaxies.

To interpret the trends observed in Figure \ref{grrigas}, we compared
the colors of the ES/SDSS sample galaxies to theoretical predictions
by running a series of Bruzual \& Charlot (2003) population synthesis
models at various metallicities and star formation histories.  The
models use the Chabrier (2003) initial mass function (IMF) to model
the stellar populations with a range of stellar masses of
0.6M$_{\odot} \leq m \leq 120{\rm M}_{\odot}$, at 6 different
metallicities (0.005, 0.02, 0.2, 0.4, 1 and 2.5 times solar) and 10
different star formation histories (SFH).  The SFHs are modeled as
having either (1) an exponentially decreasing star formation rate
(SFR; $e^{-t/\tau}$) with $\tau$ values of 8, 4, 2, and 1 Gyr (these
models have mean, mass-weighted stellar ages of 7.4, 8.6, 10.0, and
11.0 Gyr respectively); (2) an exponentially increasing SFR with
$\tau$ values of 8, 4, 2, and 1 Gyr (mean stellar ages of 4.6, 3.4,
2.0 and 1.0 Gyr respectively); (3) a continuous SFR ($\tau=\infty$;
mean stellar age of 6 Gyr); or (4) an instantaneous event of star
formation ($\tau=0$; mean stellar age of 12 Gyr).  For each model we
assume that 12 Gyr has elapsed since the first formation of stars (Dalcanton \& Bernstein 2002; Brooks et al. 2009) and
extract the SDSS model colors at that epoch.  We do not include dust
in any of the population models because we have corrected for the
internal extinction of the galaxies using the measured rotation
velocities (HI line widths) and the prescription of Tully et
al. (1998), where galaxies with the same rotation velocity are assumed
to have similar internal dust extinction. The uncertainty in this
correction is $\sim$ 0.1 magnitudes and may increase the scatter of
our photometry. However, many of the galaxies in the SDSS/ES sample
have little or no dust correction due to their slow rotation (Tully et
al. 1998; Dalcanton et al. 2004; see W09 for more details). 

Each Bruzual \& Charlot (2003) model was run without gas recycling --
the population of stars maintains the same metallicity throughout the
star formation history.  While this does not accurately mimic the
chemical evolution of individual galaxies (which are some amalgamation
of ages and metallicities), it does provide a set of specific model
comparisons by which to compare the mean properties of observed
systems.

\begin{figure}
\centering
\plotone{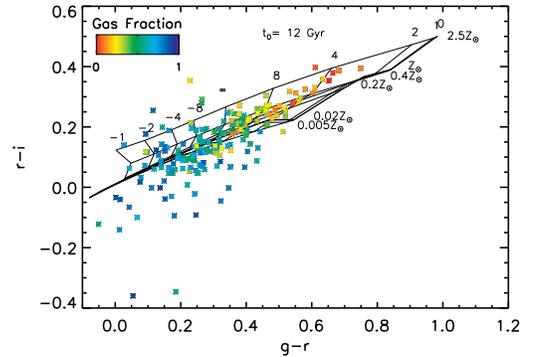}
\caption{$r-i$ vs $g-r$ for HI selected galaxies. Galaxies have been color
coded according to their gas fractions.  Bruzual \& Charlot (2003)
population synthesis grids have been overplotted for various continuous SFH and
metallicities. The SFH are
given by $\tau$ values for an exponential SFR starting at 12 Gyr ago.
($e^{-t/\tau}$).  Metallicities are plotted in comparison to solar.}
\label{bc1}
\end{figure}

Figure \ref{bc1} shows the resulting population grid for the $r-i$
colors as a function of $g-r$.  The
``horizontal'' lines are lines of constant metallicity and the
``vertical'' lines are lines of constant $\tau$ (a proxy of mean
stellar age).  Tracks of constant age are well separated in $g-r$, making $g-r$ a good probe of stellar age.  On the other hand, the $r-i$ color is
only marginally sensitive to metallicity and the aforementioned
degeneracy is apparent at all sub-solar metallicities.  These grids
bracket all reasonable continuous star formation histories, but
bracket only a narrow region of $g-r$ vs. $r-i$ color space.  Although SDSS provides photometry in 5 different photometric passbands, most of our color analysis uses only the $g$, $r$ and $i$-bands. For most of our galaxies, the $u$ and $z$ bands yield photometry with large uncertainties that does not help us differentiate between models or constrain photometric relationships.  Because the $u$-band is sensitive to the star formation rate (Hopkins et al. 2003), we have included $u$-band photometry in a limited amount of our analysis.

The ES/SDSS sample galaxies have been overplotted and color-coded
according to their gas fractions, as in Figure 1. At the red end of
the galaxy distribution, the colors are well explained by the model
grids.  The redder galaxies are consistent with having SFHs that have
been declining to the present.  These results are in agreement with
previous work by Bell et al.\ (2003) as well as Juneau et al.~(2005),
who demonstrate that galaxies with large stellar masses formed most of
their star in the first 3 Gyr after their formation.  The reddest
galaxies also have colors indicative of super-solar metallicities.
These red ($g-r >$ 0.6), high-metallicity galaxies are all early-type spiral, lenticular or elliptical galaxies.

The data and models in Figure \ref{bc1} suggest that the red galaxies
formed almost all of their stars in the distant past and have been
undergoing little to no star formation since, in spite of containing
HI at the present.  Some mechanism is therefore responsible for the
continued presence of HI.  The cannibalization of gas rich dwarfs or
infall from ``cold mode'' accretion (Katz et al.~2003; Brooks et
al.~2009) might be sufficient to explain the current small reservoir
of gas, which is only a small fraction of the galaxies' baryonic mass.
Regardless of the physical process by which these galaxies acquired or
maintained gas, the presence of HI suggests that recent star formation
is likely to have taken place.  Indeed, all of the galaxies redder
than $g-r$ $>$ 0.5 have SDSS spectra that contain emission lines. In
spite of the aperture bias in SDSS fiber spectroscopy (fibers are only
3$^{\prime\prime}$ in diameter), these spectra can be useful for
determining gross spectroscopic properties .Based on their location on
a Baldwin-Phillips-Terlevich diagram (Baldwin, Phillips \& Terlevich
1981), five of the 156 emission-line galaxies are consistent with
hosting an AGN.  All five of the galaxies with AGN have PetroR90 radii (radii containing 90\% of the Petrosian flux) larger than 30$^{\prime\prime}$.  Therefore, their disk emission (not the AGN) dominates the integrated photometry and they are included in remainder of the analysis. The majority of the galaxies in the ES/SDSS sample have emission lines that
suggest that current star formation is underway in these galaxies.

As we look to the bluer galaxies in Figure \ref{bc1}, the
inferred metallicities are lower and the SFHs move toward a constant star
formation rate ($\tau=\infty$).  This shift is likely due to a decrease in
star formation efficiency.  We define star formation efficiency as the efficiency in converting gas into stars.  As discussed above, we neglect the role of molecular gas (the state of gas from which stars form),  which should be irrelevant for the bluer galaxies in our sample. Recent studies (that included the molecular component) have seen a similar decrease in star formation efficiency in nearby, HI-dominated galaxies (Bigiel et al. 2008).  Although we do not resolve the HI emission, the failure to convert large amounts of HI into stars in the blue galaxies constitutes a decrease in star forming efficiency.  Such a decrease slows the rate of star
formation, keeping metallicities low and gas fractions high, allowing
stars to form at the present day.

The bluest galaxies ($g-r <$ 0.3) are best modeled by SFHs that have
increasing SFRs.  This area of the model parameter space has been
ignored by previous galaxy studies (Kauffmann et al.~2003; Brinchmann
et al.~2004; Salim et al.~2005; Johnson et al.~2007).  Gas infall
might serve to increase the HI surface density and subsequently the
SFR in these galaxies.  Although recent bursts have likely occurred in
these galaxies, the already blue $g-r$ colors of these do not change
significantly when recent bursts are added to the models (see \S 3.3).

Many of the bluest galaxies in the ES/SDSS sample have colors that
are not consistent with any of the Bruzual \& Charlot models.  Most of these
galaxies have $r-i$ colors that are too blue to be explained by any
iteration of the models. The LSB nature of many of these systems
suggests that for even small SFR, emission lines might be able to
dominate the broadband colors (Zackrisson et al.~2001; Magris, Binette
\& Bruzual 2003; Anders \& Alvensleben 2003; Zackrisson,
Bervall \& {\" O}stlin 2005; Zackrisson, Bergvall, \& Leitet 2008).  We investigate this possibility in detail in \S3.3.

Finally, there are a five galaxies that seem to be too red in
 $r-i$ to be explained by any population models.  Two of these
 galaxies have a saturated star nearby that is most likely affecting
 their colors.  One of the galaxies has very large photometric
 uncertainties because of its low surface brightness and the other two
 have very compact cores, which suggests the presence of AGN.  If we
 compute the colors for these two galaxies excluding the central
 region, using a shell with radial boundaries defined to be from half
 of the Petrosian radius to the Petrosian radius, both galaxies fall
 back to or below the model grids when their central core is excluded.

\subsection{The Colors of HI vs. Optically Selected Galaxies} 

\begin{figure}
\centering
\plotone{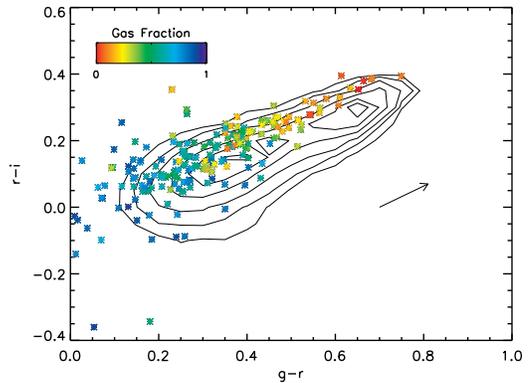}
\caption{$r-i$ vs $g-r$ for HI selected galaxies (colored symbols)
plotted with a volume selected sample of galaxies from SDSS (contours).  No internal extinction has been applied to these data but a
reddening vector has been plotted for reference.  Galaxies in the
thesis sample have been color coded according to their gas
fractions. Note that the HI selected sample appears to have colors
that are shifted off of the SDSS galaxy locus. }
\label{grri_main}
\end{figure}

Although the trends seen in Figure \ref{grrigas} have been noted in
optically selected SDSS samples (e.g. Strateva et al.~2001; Blanton et
al.~2003a), the HI selection introduces a systematic shift in the
color distribution of the ES/SDSS sample galaxies.  Figure
\ref{grri_main} shows the $r-i$ vs. $g-r$ colors for the ES/SDSS
sample plotted on top of a volume limited sample drawn from the SDSS
DR4 sample (Adelman-McCarthy et al.~2006). All galaxies have been
k-corrected and corrected for Milky Way extinction.  We used the Milky
Way extinction values from Schlegel et al. (1998) and calculated
K-corrections using the IDL package {\tt kcorrect\_v3.2} (Blanton et
al. 2003b).  The Blanton et al. (2003b) k-correction method uses
empirically-derived eigenspectra to generate model spectra from SDSS
photometry.  All of the galaxies (both SDSS DR4 and ES/SDSS) were corrected
using the same method to $z$=0.  Most of the k-corrections were smaller than 0.01 magnitudes in all bands.  No data in Figure
\ref{grri_main} have been corrected for internal extinction.  The
reddening vector (derived from Schlegel et al.~1998) has been included
to show that any correction to either data set is almost perfectly
aligned with the galaxy locus.

\begin{figure}
\centering
\plotone{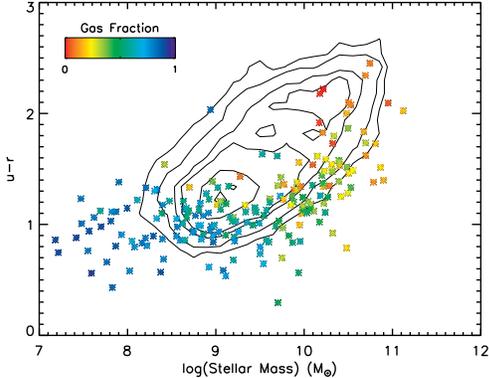}
\caption{The $u-r$ colors as a function of stellar mass for the ES/SDSS galaxies (colored according to gas fraction) and the SDSS DR4 sample (contours).  The bimodal galaxy distribution (blue and red sequence) is easily seen in the SDSS DR4 galaxies, but absent from the ES/SDSS galaxies.  The HI selection of the ES/SDSS galaxies identifies gas-rich galaxies that are on the blue sequence and are currently in the process of star formation.}
\label{smassur}
\end{figure}

The HI-selected galaxies do not show the color bimodality seen in previous SDSS studies (e.g. Blanton et al.~2003a; Baldry et al.~2004).  Figure \ref{smassur} shows the $u-r$ colors as a function of stellar mass for the ES/SDSS galaxies (colored according to gas fraction) and the SDSS DR4 sample (contours).  The bimodal galaxy distribution (blue and red sequence) is easily seen in the SDSS DR4 galaxies, but is absent from the ES/SDSS galaxies.  The HI selection of the ES/SDSS galaxies identifies gas-rich galaxies that are on the blue sequence and are currently in the process of star formation.  Like in Figure \ref{grri_main}, the ES/SDSS galaxies in Figure \ref{smassur} appear to be offset from the SDSS DR4 sample (bluer in $g-r$ or redder in $r-i$).  Figure \ref{smassur} also demonstrates the ability of an HI-selected sample to recover galaxies with small stellar masses; the ES/SDSS sample has a significant number of galaxies with stellar masses below 10$^8$ M$_{\odot}$.

It is clear from Figure \ref{grri_main} that the HI selected sources
fall to one side of the SDSS galaxy distribution, such that they are
either bluer in $g-r$ or redder in $r-i$. This offset is particularly
pronounced at the red end of the color distribution where the color
dispersion is small.  As noted in the previous section, the redder
galaxies have SFHs that suggest that almost all star formation occurred
in the distant past.  The presence of HI in these systems may indicate
a recent acquisition of HI and a subsequent ignition of star
formation, as confirmed by their SDSS spectra.  This temporary increase
in SFR due to late-time accretion would naturally explain why the red HI
selected galaxies are slightly bluer in $g-r$ than the rest of the
SDSS galaxy population (Figure \ref{grri_main}).

To constrain the amount of recent star formation needed to produce the
observed color offset between the ES/SDSS galaxies and the ``main'' SDSS
sample, we re-ran the same series of Bruzual \& Charlot (2003)
population synthesis codes as in Figure \ref{bc1}, but with the
addition of a recent burst of star formation.  We generated an
initial model with $g-r$ and $r-i$ colors of 0.75 and 0.3
respectively ($\tau$=2; Z=Z$_{\odot}$). We
then added a single instantaneous burst (delta-peak) of star formation to the continuous model
and varied the time of the burst from 2 Gyr to 100
Myr in the past.  We also varied the burst strength from 0.1-10\% of the
integrated star formation.  The resulting models were most sensitive
to the time at which the burst was placed. The models that most
closely matched the slight $g-r$ bluing were bursts placed at 300 Myr
in the past with 1\% of the past integrated star formation occurring in
the burst. 

\begin{figure}
\centering
\plotone{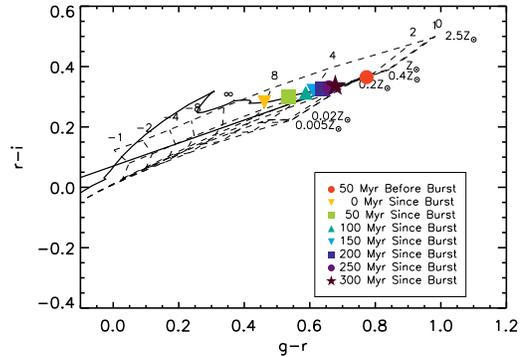}
\caption{The $r-i$ vs. $g-r$ evolution of a single Bruzual \& Charlot
  model with the addition of an instantaneous burst of star formation
  that produced 1\% of the total stellar mass of the galaxy and
  occurred 300 Myr in the past (solid line). The Bruzual \& Charlot
  continuous star formation grid (from Figure \ref{bc1}) is plotted
  for reference (dashed lines). The continuous model to which the
  burst was added corresponds to a red galaxy from the ES/SDSS sample
  ($\tau$=2, Z=Z$_{\odot}$; $g-r$ and $r-i$ colors of 0.75 and 0.3
  respectively). Colored symbols are plotted every 50 Myr starting 50
  Myr before the burst. Colors bluer than $g-r$ $<$ 0.4 and $r-i$ $<$
  0.2 are observed only for very short durations ($<$1 Myr).  By 100
  Myr the colors have reddened to within $\Delta g-r$=0.15 and $\Delta
  r-i$=0.05 of their initial value and continue to redden for hundreds
  of Myr.}
\label{bursttime}
\end{figure}

Figure \ref{bursttime} shows the evolutionary track for the best
matched model burst (300 Myr in the past) with symbols plotted every
50 Myr since burst. The figure demonstrates that there is a rapid and
short-lived change in color that occurs in the first few Myrs and is
followed by a slow reddening for hundreds of Myrs.  Colors bluer than
$g-r$ $<$ 0.4 and $r-i$ $<$ 0.2 are observed only for very short
durations ($<$1 Myr).  By 100 Myr the colors have reddened to within
$\Delta g-r$=0.15 and $\Delta r-i$=0.05 of their initial value and
continue to redden for hundreds of Myr.  This ``bottleneck'' in the
burst evolution suggests that the small color offset observed in
Figure \ref{grri_main} could be sufficiently long lived to be due to a
burst in the past few hundred Myr.

\begin{figure}
\centering
\plotone{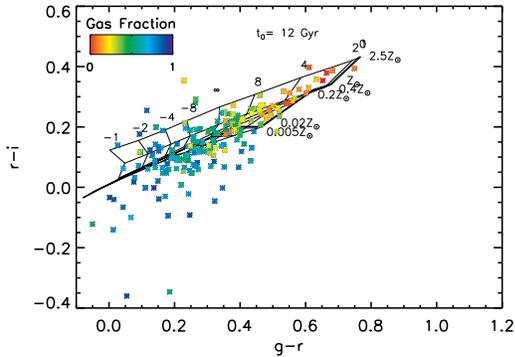}
\caption{$r-i$ vs $g-r$ for HI selected galaxies.  All photometric
  corrections have been applied to these data. Galaxies have been
  color coded according to their gas fractions.  Bruzual \& Charlot
  (2003) population synthesis grids have been overplotted for various
  SFH and metallicities after 12 Gyrs of star formation. The SFH are
  given by $\tau$ values for an exponential SFR ($e^{-t/\tau}$).
  These grids include an instantaneous burst of star formation that
  produced 1\% of the stellar mass, 300 Myr in the past.}
\label{bc2}
\end{figure}

Figure \ref{bc2} shows the ES/SDSS galaxies and the model grids for
continuous star formation with an additional 1\% burst occurring
300 Myr in the past.  The red HI selected galaxies in the burst model
have $\tau \sim3$, compared to the $\tau \sim4$ seen in the burst-free
models.  By comparing the difference between the grids in Figure
\ref{bc1} and \ref{bc2} with the offset observed in Figure
\ref{grri_main}, it is clear that the HI selected galaxies in the
burst case (Figure \ref{bc2}) occupy the same model space ($\tau
\sim3$) as the SDSS ``main'' galaxies in the burst-free case (Figure
\ref{bc1}).  Unfortunately, our inability to correct for internal
extinction in the SDSS ``main'' sample prohibits us from overlaying
accurate grids on Figure \ref{grri_main}, which would be highly
instructive.  However, ``blinking'' back and forth between Figures
\ref{bc1} and \ref{bc2} confirms that a recent burst can explain
the offset of the red galaxies.

It is important to note that the addition of the burst may effect the
colors of the blue galaxies as well.  A single instantaneous burst
produces extremely blue colors for very short times. However, in
reality, the bursts are not instantaneous, but instead last for a
finite amount of time with some exponential decrease (10-300 Myr
depending on galaxy type; McQuinn et al. 2009).  If a significant
fraction of the star formation has occurred in the very recent past,
then some of the blue colors may be due to recent bursts. However,
rising SFRs are still required to explain the bulk of the galaxies
with colors bluer than $g-r <$0.3.  

 Although degeneracies between the time, duration and strength of the
 star formation events in the models prevent an exact characterization
 of the bursts in the HI selected galaxies, we can safely claim that a
 new generation of stars has been formed in the last few hundred Myr
 in the red systems (a large instantaneous burst has a similar
effect on the integrated colors as a small exponentially decreasing
burst). The bursts require no more than 1\% of the mass
 to be involved as long as the burst was in the last 300 Myr.

 We note that the bluer $g-r$ colors in the red ES/SDSS galaxies may not be solely due to main sequence stars, since most blue main sequence stars will have ceased their hydrogen burning after a few hundred Myr; the $g-r$ color of a main sequence turnoff star at 300 Myr is $\sim$0.0 (Sarajedini et al.~2004; Covey et al.~2007). We suggest that another possibility for the blue
 $g-r$ colors we see in the red HI selected galaxies could be due
 to an elevated level of BHeB stars left over from a recent burst of
 star formation. These stars are seen several hundred Myr after a star
 formation event (Dohm-Palmer et al.~2002).

\subsection{Modeling Galaxy Colors with Emission Lines}

Even with the addition of bursts, the colors of many ES/SDSS galaxies
do not fall on the Bruzual \& Charlot (2003) model grids.  Unlike the
red galaxies, which can be easily explained, the majority of the
discrepant galaxies are significantly bluer in $r-i$ than any models
can predict.  These systems also have high gas fractions and most have
low surface brightness. No matter what parameters we alter in the
Bruzual \& Charlot models, we cannot produce a stellar population with
very blue $r-i$ colors.

We show that the discrepant colors can be explained by including the
emission lines that dominate the spectra of these outlying galaxies.
The blue colors and high gas fractions of the discrepant galaxies
suggest that they are actively forming stars.  Because these
galaxies also have low surface brightnesses, it is possible that
emission lines produced by HII regions contribute significantly to
their luminosity.  HII regions produce numerous emission lines in the
$g$ and the $r$ bands. H$\alpha$ dominates in the $r$-band while [OIII]
and H$\beta$ dominate in the $g$-band.  In contrast, the $i$-band has almost no
emission lines.  Therefore, in an HII region, we would expect the
$r-i$ color to be significantly bluer than the underlying stellar
continuum and the $g-r$ color to change only slightly. This suggests that $i$-band luminosity may be a better indicator of the underlying stellar mass than $g-r$. It also implies that the $g-r$ color should be used to compute color dependent mass-to-luminosity ratios (the emission lines tend to cancel out; see below). 

We can test this idea using the SDSS spectroscopic fibers, which were
often placed on HII regions for LSB galaxies.  By comparing the colors
of HII regions (derived from SDSS spectra) with the integrated colors
of the galaxies that host them, we can quantify color changes produced
by HII regions.

\begin{figure}
\centering
\plotone{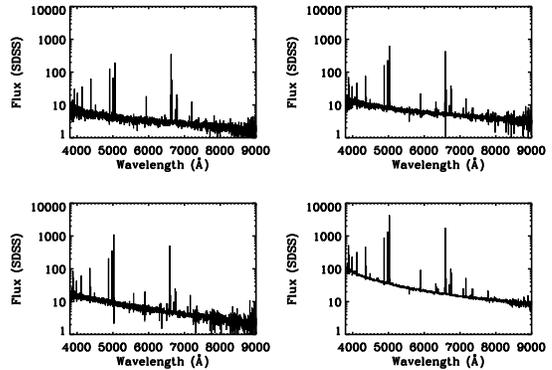}
\caption{SDSS fiber spectra for 4
galaxies in the ES/SDSS sample with strong emission line features.
All of the fibers for these spectra were placed on top of HII regions
in the galaxies.  The flux axis is in units of 10$^{-17}$
ergs/s/cm$^2$\AA.}
\label{spectra}
\end{figure}

We first selected 10 ES/SDSS galaxies with SDSS spectroscopy and
required that they have large H$\alpha$ equivalent widths ($>$ 300
\AA).  Four of the spectra selected are shown in Figure \ref{spectra}.
All of the selected spectra have good signal-to-noise ($>$ 50) and are
dominated by emission line features.  We visually inspected each
galaxy to ensure that all of the spectra selected (and shown in Figure
\ref{spectra}) are from fibers placed on high surface brightness HII
regions.  We convolved the $g$, $r$ and $i$ SDSS filter curves with
each spectrum and converted the computed flux density to AB magnitudes
using the relation that a magnitude 0 object has a flux density of
3631 Jy (Oke \& Gunn 1983). We then compared the integrated colors of
the 10 galaxies derived in W09 with the HII region colors derived from
the spectra.

\begin{figure}
\centering
\plotone{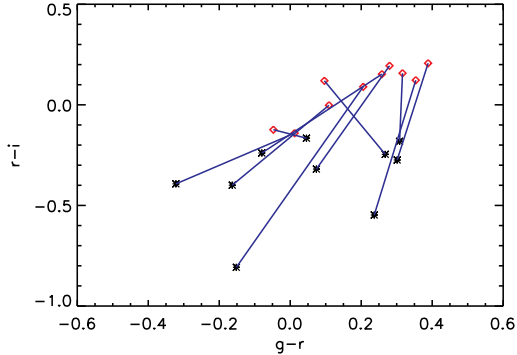}
\caption{$r-i$ vs. $g-r$
color difference between the integrated colors of 10 galaxies (red
diamonds) and the ``fiber'' magnitudes computed from the HII region
spectra (black symbols).  The blue lines connect each galaxy's
integrated color to its HII region color.  All of the galaxies get
bluer in $r-i$ and all but 2 get bluer in $g-r$}
\label{em1}
\end{figure}

Figure \ref{em1} shows the resulting differences between the HII
region and the integrated color of each galaxy.  The colors derived
from fibers placed on HII regions are plotted as black symbols and the
integrated galaxy colors derived from images (W09) are plotted as
red diamonds.  Blue lines
connect a galaxy's integrated color to the color of the
3$^{\prime\prime}$ diameter aperture around the HII regions.  The HII region
colors are bluer in $r-i$ by $\sim$0.5$^{m}$ and all but 2 become bluer
in $g-r$ by $\sim$0.2$^{m}$, compared to the integrated colors.  This result
confirms the hypothesis that emission lines can make the $r-i$ color
blue enough to explain the colors we see in the ES/SDSS sample.
However, contrary to our initial expectation, the $g-r$ color does change
considerably.  The bluer $g-r$ colors near HII regions can be
explained as being produced by a young underlying stellar
population. More surprisingly, the colors near two of HII regions are
redder in $g-r$ than the integrated light from the galaxies.  This
difference may reflect the presence of dust localized to the HII region.

To further test the emission line hypothesis as well as to confirm the
reason for the $g-r$ color change in Figure \ref{em1}, we added model emission lines to
the computed Bruzual \& Charlot model colors.  Theoretical models for
line emission in galaxies are given by Kewley et al.~(2001) using the
{\tt STARBURST99-MAPPINGS III} code.  The models cover a range of
metallicities, ionization states, electron densities, and assumption
about the burst lengths.  For each model, the wavelength of every line
and its energy are provided.  We convolved these energies with the
SDSS filters and converted the computed flux densities into AB
magnitudes.  We added the emission line output to the Bruzual \&
Charlot colors using a scaling coefficient that determines the
strength of the emission line relative to the true stellar
continuum. This coefficient is directly proportional to the SFR and
can be used to estimate the range of possible SFRs for a galaxy.  The
final models are not fully self consistent, because the ionizing
radiation for the emission line spectrum does not come directly from the
Bruzual \& Charlot spectrum.  However, because the star formation is
highly localized in late-type galaxies, it is possible for the
ionizing radiation in isolated HII regions to be different than that
of the broader component of stars

When adding the {\tt STARBURST99-MAPPINGS III} emission-line models to
the Bruzual \& Charlot outputs, we ensured that the adopted gas
metallicity matched the assumed stellar metallicity.  Because there
are only 3 overlapping metallicities, we limited our analysis to 0.2
Z$_{\odot}$, 0.4 Z$_{\odot}$ and solar metallicity populations.  In reality most
of the galaxies where emission lines dominate are low mass and gas
rich and thus are unlikely to have high metallicities.  Even values of
0.2 Z$_{\odot}$ might be too high for these systems (Tremonti et
al.~2004).

\begin{figure}
\centering
\plotone{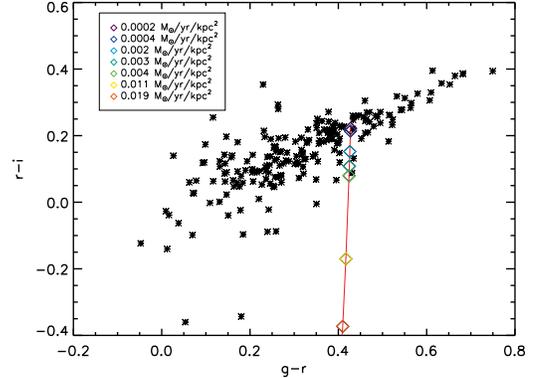}
\caption{The $r-i$ vs. $g-r$ colors of the ES/SDSS galaxies with the addition of
  emission lines to an underlying stellar population for a single
  Bruzual \& Charlot model galaxy ($\tau$=8 Gyr; Z=0.4 Z$_{\odot}$). The
  colored symbols represent different star formation rates. The
  galaxy's $r-i$ color becomes bluer as star formation is increased,
  while the $g-r$ color goes almost unchanged.}
\label{emissionsim}
\end{figure}

\begin{figure}
\centering
\plotone{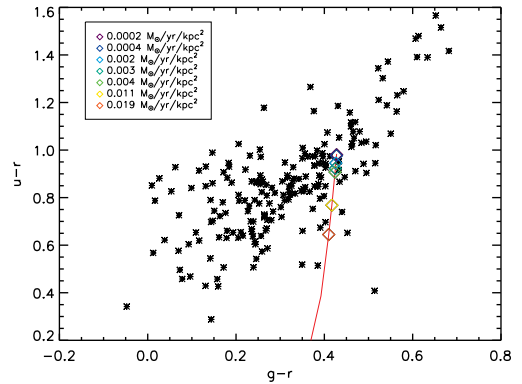}
\caption{The $u-r$ vs. $g-r$ colors of the ES/SDSS galaxies with the addition of  emission lines to an underlying stellar population for a single
  Bruzual \& Charlot model galaxy ($\tau$=8 Gyr; Z=0.4 Z$_{\odot}$). The
  colored symbols represent different star formation rates.}
\label{emissionsim2}
\end{figure}

Figures \ref{emissionsim} and \ref{emissionsim2} show the colors
($r-i$ vs.$g-r$ and $u-r$ vs. $g-r$ respectively) that result from
adding emission lines to a single Bruzual \& Charlot model ($\tau$=8
Gyr; Z=0.4Z$_{\odot}$). This particular model was chosen because it lies on the red end of the blue population and intersects the ES/SDSS color-color locus.  The colored symbols represent different star
formation rates.  In Figure \ref{emissionsim}, the model galaxy's
$r-i$ color becomes bluer as star formation is increased, while the
$g-r$ color is almost unchanged. A similar effect is seen in the $u-r$
color in Figure \ref{emissionsim2}.  This confirms that the non-zero
$g-r$ color offsets near HII regions must indeed be due to stellar
populations and or dust.

\begin{figure}
\centering
\plotone{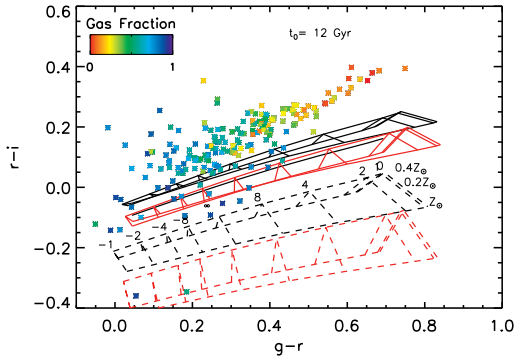}
\caption{$r-i$ vs. $g-r$ colors for the HI selected
  galaxies.  Color coding indicates the gas fractions of the
  galaxies. The grids represent 2 different linear combinations of the
  STARBURST99-MAPPINGS III emission line models and the Bruzual \&
  Charlot population synthesis models.  The Bruzual \& Charlot models
  are identical to those shown in Figure \ref{bc1}.  All emission line
  models use an 8 Myr continuous burst of star formation.  Black lines
  indicate emission line models with an electron density of 10
  cm$^{-3}$ and an ionization parameter of 5$\times10^6$ cm\ s$^{-1}$.  Red
  lines have the same electron density but an ionization parameter of
  8$\times10^7$ cm\ s$^{-1}$.  Solid lines have SFR densities of
  1.2$\times10^{-3}$ M$_\odot$\ yr$^{-1}$\ kpc$^{-2}$ and 3.7$\times10^{-3}$
  M$_\odot$\ yr$^{-1}$\ kpc$^{-2}$ for the two models (black and red)
  respectively.  The dashed lines have SFRs 3 times larger than the
  solid lines.  These models clearly describe the colors of all of the
  galaxies that are not fit by normal populations of stars.}
\label{sb99grids}
\end{figure}

In Figure \ref{sb99grids}, we show stellar population grids similar to those in Figure \ref{bc1}, but with an 8 Myr long continuous burst of star formation.  Black lines indicate emission line models with an electron density of 10 cm$^{-3}$ and an ionization parameter of 5$\times10^6$ cm~s$^{-1}$.  Red lines have the same electron density but an ionization parameter of 8$\times10^7$ cm~s$^{-1}$. This plot demonstrates that adding emission lines of
varying strength can explain all of the discrepant blue galaxies.  While 8 Myr may be short for some realistic starburst events, it successfully demonstrates the effect of current (or recent) star formation on the colors of galaxies.

We found that varying the electron density and burst length (8 Myr
continuous vs. instantaneous) did not have a major effect on the
output models. All significant changes in color are driven by the
ionization parameter and by the global normalization of the emission
line spectrum, which is directly proportional to the SFR (the emission
lines are generated for a specific value of the SFR).  Solid lines in
Figure \ref{sb99grids} have SFR densities of 1.2$\times10^{-3}$
M$_\odot$~yr$^{-1}$~kpc$^{-2}$ and 3.7$\times10^{-3}$
M$_\odot$~yr$^{-1}$\ kpc$^{-2}$ for the two models (black and red)
respectively.  The dashed lines have SFRs 3 times larger than the
solid lines.  For typical galaxies in our sample, we estimate SFRs of
0.004-0.2 M$_\odot$~yr$^{-1}$ in the emission line dominated regions.
These are reasonable SFRs for small, blue galaxies (van Zee 2001;
Hunter \& Elmegreen 2004; Salzer et al.~2005).  We estimated the SFR
density of the ES/SDSS sample using the $u$-band luminosity as a proxy
for SFR (Hopkins et al.~2003).  We found that the SFR densities range
from 0.0001 to 0.02 M$_\odot$~yr$^{-1}$~kpc$^{-2}$.  This range is
consistent with the star formation rate densities required to produce
the blue offset in the galaxies in Figure \ref{emissionsim}.  While
the models might not be perfectly self-consistent as discussed above,
they do demonstrate that strong emission lines can certainly affect
the colors of galaxies.  We note that the ``pitched roof'' shaped grids are a result of how the emission lines are affected by slight changes in the metallicity.  It is clear that the span of these models can
easily explain the galaxies that the stellar population models cannot
and that many low mass, blue galaxies must have colors that are
dominated by emission lines (e.g. Zackrisson et al.~2005).

The emission line effect may also amplify the underlying color offset discussed in the previous section by making galaxies appear bluer in $r-i$ than they actually are.  However, the colors of the red galaxies in the offset region are dominated by their stellar populations and not likely affected by the emission line colors.

\subsection{The Increase in Color Dispersion}

The analysis above demonstrates that the observed colors of the
ES/SDSS sample can be well explained by stellar population models
including emission lines, and that the general trend in color is due
to decreasing mean stellar age with increasing gas fraction. However,
while this explains most of the general trends seen in the color
distribution in Figure \ref{grrigas}, it does not fully explain the increase in the dispersion of color from red to blue galaxies.  The narrow
dispersion at red colors grows to a much larger value before the
photometric uncertainties become large enough to be solely responsible
for the discrepancy.  Although some of the increased dispersion is likely due to the onset of emission line dominated colors, this cannot account for the entirety of the color dispersion.

To explore how the dispersion in color relates to other physical
quantities, we performed a principal component analysis (PCA) on the
$r-i$ vs. $g-r$ colors.  The result aligns the principal axis (P1)
with the galaxy color-color locus and sets the secondary axis (P2)
perpendicular to P1.  P1 and P2 can be expressed in term of the $g-r$
and $r-i$ colors of the galaxies by

\begin{equation}
P1= 0.6(r-i)+0.8(g-r)-\langle 0.6(r-i)+0.8(g-r)\rangle
\end{equation}

\noindent and

\begin{equation}
P2=0.8(r-i)-0.6(g-r)-\langle 0.8(r-i)-0.6(g-r)\rangle,
\end{equation}

\noindent where the angle brackets indicate the mean of the enclosed quantity for the sample (the resulting eigenvalues are 6.2 and 0.8 for the eigenvectors pointing along the P1 and P2 axis respectively).

\begin{figure}
\centering
\plotone{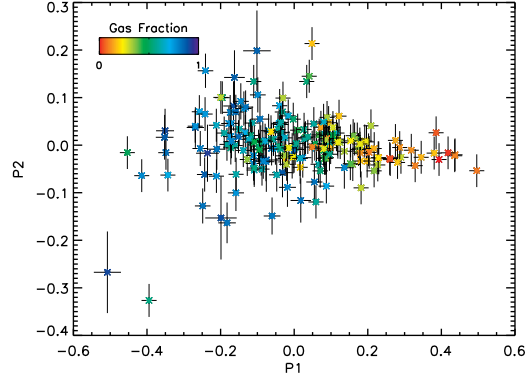}
\caption{Principal axis of the galaxy color-color relation after PCA.  Galaxies are color coded according to their gas fraction.  Error bars were calculated using the uncertainties in the $g$, $r$ and $i$-band photometry.}
\label{pca}
\end{figure}

Figure \ref{pca} shows the result of the PCA coordinate
transformation.  Error bars were calculated using the uncertainties in
the $g$, $r$ and $i$-band photometry.  The P2 axis serves as an
indicator of dispersion from the galaxy locus and we compare the P2
values to other physical parameters below.

\begin{figure}
\centering
\plotone{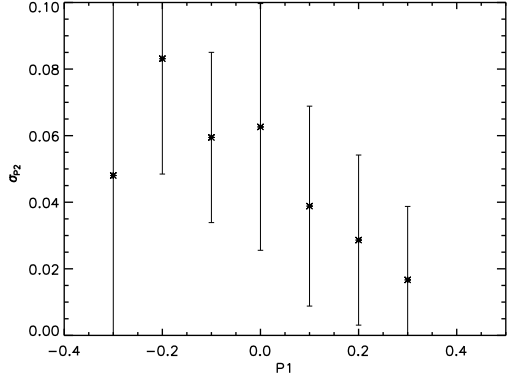}
\caption{The rms scatter in the P2 axis as a function of P1.  The error bars indicate the combined photometric error of the data points in each bin.}
\label{p1p2}
\end{figure}

We quantified the dispersion in the P2 axis by calculating the rms P2 scatter in bins of P1. Figure \ref{p1p2} shows the resulting change in color dispersion along the principal axis.  The error bars indicate the photometric uncertainty in the P2 value.  As seen in Figures \ref{pca} and \ref{grrigas}, blue galaxies have a significantly larger color dispersion than red galaxies.

Changes in observable properties of the galaxies, such as stellar
mass, rotation velocity, surface density and surface brightness might be related to a change in their color dispersion.  As seen in previous work on the
ES/SDSS sample, many of these physical properties are related to
each other (Disney et al.~2008; Garcia-Appadoo et al.~2009) and a
change in one property is likely to show up as a variation in the
others.

\begin{figure}
\centering
\plotone{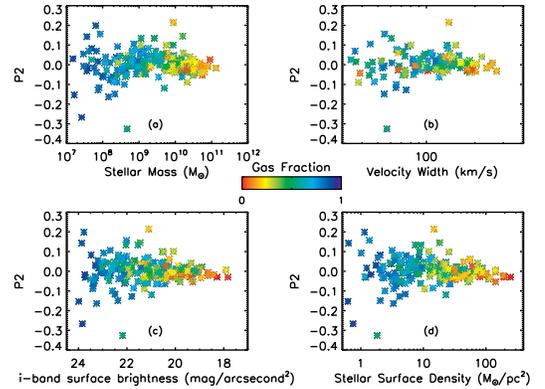}
\caption{P2 PCA axis as a function of stellar mass (a), velocity width (b), $i$-band surface brightness (c) and stellar surface density (d).  Points have been color coded according to their gas fractions.}
\label{varp2}
\end{figure}

\begin{figure}
\centering
\plotone{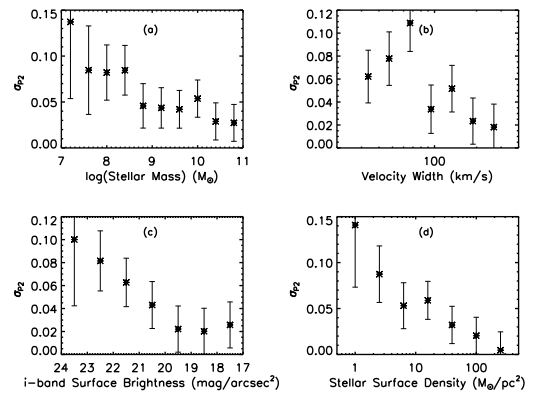}
\caption{The rms scatter in the P2 axis as a function of stellar mass (a), velocity width (b), $i$-band surface brightness (c) and stellar surface density (d).  The error bars represent the combined photometric error in each bin.}
\label{sigp2}
\end{figure}

To examine how the change in color dispersion relates to the physical
properties of the ES/SDSS galaxies, we plotted the P2 PCA values as a
function of stellar mass, rotation velocity (measured from half of the
20\% HI velocity width), surface brightness and stellar surface
density (calculated by measuring the average stellar mass density
within the 90\% Petrosian radius). Figure \ref{varp2} shows that
changes in the physical properties trace the change in dispersion
very well.  In addition, Figure \ref{varp2} has been color coded
according to gas fraction.  As seen in previous studies, the gas
fractions also correlate with the physical properties of the ES/SDSS
galaxies (Garcia-Appadoo et al.~2009).  We quantified the change in
color dispersion by plotting the rms scatter in the P2 axis as a
function of the four physical properties (Figure \ref{sigp2}).  All
four properties correlate with the rms scatter in the P2 variable.  We
computed the Spearman's rank correlation coefficient for each relation
to test which property best correlates with the color dispersion.  We
calculated Spearman's $\rho$ values of 0.92, 0.82, 0.89 and 0.96 for
the stellar mass, velocity, surface brightness and surface density
respectively.  The stellar density shows the best correlation with
color dispersion with the lowest density galaxies having the highest
color dispersion.

The increase in color dispersion as a function of surface brightness
and stellar surface density may probe a change in the star formation
process.  At the low surface brightnesses, the stochastic nature of
star formation (Stinson et al.~2007), and the dominance of emission
lines in low-mass galaxies, controls the observed colors and creates a
larger color dispersion, while the colors of higher surface brightness
systems are dominated by stellar populations (and to a smaller extent
recent bursts of star formation, as discussed above).  Because the
surface brightness and stellar surface density are tied closely to other properties of the ES/SDSS galaxies (Disney et al.~2008), this change in color dispersion may be evidence for a significant change in the way stars form in
galaxies.

While the stellar mass, surface brightness and stellar surface density change smoothly, the velocity width shows a statistically significant (4 $\sigma$) jump at a velocity of 80 km~s$^{-1}$.  Previous studies have found that the  rotation velocity in galaxies correlates with transitions in their disk properties.  Dalcanton, Yoachim \& Bernstein (2004) showed that the presence of dust lanes in a sample of edge-on disk galaxies is related to rotation
speed.  At $V_c < 120$ km~s$^{-1}$, no dust lanes are present in their
galaxies.  Disk instabilities occurring at $V_c > 120$ km~s$^{-1}$ act to
collapse the dust to a small scale height and produce a larger optical
depth.  Star formation efficiency is also increased above this
velocity threshold, affecting the SFH and colors of the faster rotators
(Verde et al.~2002; Dalcanton et al.~2004).   Lee et al.~(2007) find an additional transition around 50 km~s$^{-1}$, where the dispersion in H$\alpha$ EW dramatically increases.  This transition is interpreted as a possible change from large-scale star formation in spiral galaxies to stochastic star formation in irregular systems (e.g. Stinson et al.~2007).  We observe a transition in the color dispersion at 80 km~s$^{-1}$ (Figure \ref{sigp2}b) that lies in between the Dalcanton and Lee transitions.  The transition in the ES/SDSS sample may be a result of a similar physical process, namely the change from well-organized star formation in massive disks to an irregular stochastic process in lower mass systems

\begin{figure*}
\centering
\plotone{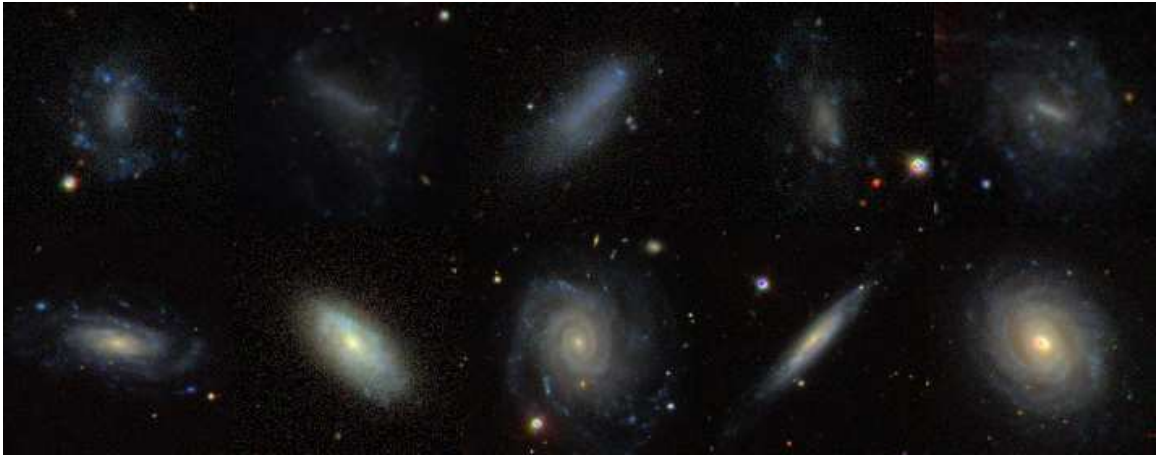}
\caption{5 randomly selected $gri$ composite images of galaxies from the slowly rotating regime (V$_{rot}$ $<$ 80 km~s$^{-1}$; top row) and 5 composite images from the faster regime (V$_{rot}$ $>$ 80 km~s$^{-1}$; bottom row). [A higher resolution version of this figure is available from ADS and the Astronomical Journal.]}
\label{locus}
\end{figure*}

To illustrate the differences between systems across the velocity
transition, we randomly selected 5 $gri$ composite images of galaxies
from the low velocity (V$_{rot}$ $<$ 80 km~s$^{-1}$) population (top row of
Figure \ref{locus}) and 5 images from the high velocity population (V$_{rot}$
$>$ 80 km~s$^{-1}$; bottom row of Figure \ref{locus}).  Figure \ref{locus} reveals a clear morphological distinction between the two populations, with the higher velocity systems having higher surface brightnesses, earlier morphological types and a more organized distribution of stars (an indication of large scale star formation).

\section{Discussion}

We used the population synthesis models of Bruzual \& Charlot (2003) to
model the SFHs and metallicities of galaxies in the ES/SDSS sample.
We found that red galaxies have super-solar metallicities and have SFHs that are consistent with them forming the bulk of their stars in the distant past.  These red galaxies have presumably exhausted their early gas supply but have recently acquired gas through mergers and infall. Although the infalling gas is likely low metallicity and could in principle dilute the metallicity, the small amount of gas in red galaxies (low gas fractions) is not enough to make an observable difference in the average metallicity of the ES/SDSS systems.

Bluer galaxies have lower metallicities and their mean stellar ages
are younger (increasing $\tau$ values).  The gas fractions suggests
that these blue systems have at least as much gas as they do
stars. Previous studies have indicated that these types of galaxies are less massive and have disks that are stable against collapse, making their
star formation much less efficient.  They have also likely retained
much of their initial HI and are slowly converting it into stars as is
suggested by the large (or negative) $\tau$ values.  As galaxies pass
the threshold for disk stability, their star formation becomes
sporadic and there is a clear dispersion in the colors due to
variations in burst age, burst strength and superimposed emission
lines.  The idealistic Bruzual and Charlot models are likely only
loose guides for these systems, as their star formation histories no
longer follow smooth exponential functions.

Figure \ref{bc1} indicates that there are a few galaxies in the regime of exponentially increasing SFR (negative $\tau$ values).  This is not surprising as there are few mechanisms that will increase the SFR of a galaxy.  The likely culprit for these systems is the infall of gas.  As gas is accreted, the gas densities in the galaxies will increase and local gravitational collapse will become more efficient.  This will increase the SFR as well as the eventual metallicity of the galaxy, pushing its $g-r$ colors to the blue (more recent star formation) and its $r-i$ colors slightly to the red (more metals).  

We also showed that HI selected galaxies are offset from the SDSS galaxy locus, especially at the red end, and that this is likely due to bursts of star formation in the past few hundred Myr.  The gas that induces these recent bursts is not primordial and is best explained by the accretion of gas rich dwarfs.  
  
The bluest galaxies in the ES/SDSS sample are not explained by population synthesis models alone.  Their colors can be modeled only with the inclusion of emission lines.  We showed that emission line spectra with reasonable SFRs can explain the colors of the bluest galaxies in our sample. 

We also showed that the distribution of galaxies at the red end of the
color-color locus has a very small dispersion that continues to rise
into the blue regime. This change in dispersion appears to correlate
with stellar mass, rotation velocity, surface brightness and
especially stellar surface density. It is possible that the change in
dispersion is evidence for a significant change in the way stars form
in galaxies at a given mass scale.  Massive galaxies have unstable
disks and efficiently convert most of their primordial reservoirs of
gas into stars in the first few Gyrs after their formation.  These
galaxies are bulge dominated and have redder, older and more massive
populations of stars.  We do see a sharp transition in the color
distribution at a rotation velocity of 80 km~s$^{-1}$.  This
transition may be related to previous results that have found sharp
transitions in galaxy properties as a function of rotational velocity
(Dalcanton et al.~2004; Lee et al.~2007).  However, further
investigations are required to explore the various velocity
transitions as a function of galaxy properties.

It may also be the case that the increase in dispersion is nothing
more than a selection effect related to the surface brightness. As the
surface brightness of a system decreases, the effect of a single burst
of stars on the color becomes increasingly large, possibly explaining
the increase in scatter.  This effect might explain the onset of
emission line dominated colors in the bluest galaxies as they can
influence of the colors of galaxies devoid of massive stellar
populations.  

We note that the change in color dispersion is not easily seen in the
volume selected SDSS data.  Applying the HI selection identifies a low
dispersion subsample of the SDSS galaxies, most notably at the red
end.  As mentioned above, the red HI selected galaxies appear to be
bluer in $g-r$ than the ``main'' SDSS sample. These two features are
likely related.  It is possible that the large distribution in color
at the red end of the SDSS main sample is due to the diversity of gas
content.  Because most of the red galaxies in SDSS exhausted their
original supply of gas long ago, this dispersion may be an indication
of the spread in time since the last major gas infall.  We leave
further discussion of the ``color offset'' to future study.
 
\section{Acknowledgments}
AAW and JJD acknowledge the support of NSF grant 0540567, the Royalty
Research Fund, and ADVANCE. AAW also acknowledges support from the
Astronaut Scholarship Foundation. JJD was partially supported by the
Wyckoff Faculty Fellowship.  Many thanks are extended to the anonymous
referee for his/her many useful suggestions for improving this
manuscript. The authors would like to thank Mike Blanton, Marla Geha,
Beth Willman, Paul Hodge, Suzanne Hawley, Tom Quinn, Peter Yoachim,
Anil Seth, and Vandana Desai for useful discussions while completing
this project.  We also thank John Bochanski for technical assistance
and Adam Burgasser for his financial support and insistence that this
project reach completion.

Funding for the Sloan Digital Sky Survey (SDSS) and SDSS-II has been
provided by the Alfred P. Sloan Foundation, the Participating
Institutions, the National Science Foundation, the U.S. Department of
Energy, the National Aeronautics and Space Administration, the
Japanese Monbukagakusho, and the Max Planck Society, and the Higher
Education Funding Council for England. The SDSS Web site is
http://www.sdss.org/.

The SDSS is managed by the Astrophysical Research Consortium (ARC) for
the Participating Institutions. The Participating Institutions are the
American Museum of Natural History, Astrophysical Institute Potsdam,
University of Basel, University of Cambridge, Case Western Reserve
University, The University of Chicago, Drexel University, Fermilab,
the Institute for Advanced Study, the Japan Participation Group, The
Johns Hopkins University, the Joint Institute for Nuclear
Astrophysics, the Kavli Institute for Particle Astrophysics and
Cosmology, the Korean Scientist Group, the Chinese Academy of Sciences
(LAMOST), Los Alamos National Laboratory, the Max-Planck-Institute for
Astronomy (MPIA), the Max-Planck-Institute for Astrophysics (MPA), New
Mexico State University, Ohio State University, University of
Pittsburgh, University of Portsmouth, Princeton University, the United
States Naval Observatory, and the University of Washington.

\bibliographystyle{apj}



\end{document}